\documentclass[amsfonts, amssymb, prl, superscriptaddress, notitlepage, twocolumn, nofootinbib]{revtex4-2}

\usepackage{amsmath}    
\usepackage{graphicx}   
\usepackage{verbatim}   
\usepackage{color}      
\usepackage{hyperref}   
\usepackage[english]{babel}
\usepackage{ucs}
\usepackage[utf8x]{inputenc}
\usepackage{subfigure}
\usepackage[autostyle]{csquotes}
\usepackage{float}
\usepackage{textcomp}
\usepackage{xspace}
\usepackage{orcidlink}

\newcommand{\Corr}{\operatorname{Corr}}

\begin{document}

\title{Two scholarly publishing cultures? Open access drives a divergence in European academic publishing practices}
\author{Leon~Kopitar \orcidlink{0000-0002-6647-9988}}
\affiliation{University of Maribor, Faculty of Health Sciences, Slovenia}
\affiliation{University of Maribor, Faculty of Electrical Engineering and Computer Science, Slovenia}

\author{Nejc~Plohl}
\affiliation{University of Maribor, Faculty of Arts, Slovenia}

\author{Mojca~Tancer Verboten}
\affiliation{University of Maribor, Faculty of Law, Slovenia}
\affiliation{University of Maribor, Faculty of Chemistry and Chemical Engineering, Slovenia}

\author{Gregor~Štiglic \orcidlink{0000-0002-0183-8679}}
\affiliation{University of Maribor, Faculty of Health Sciences, Slovenia}
\affiliation{University of Maribor, Faculty of Electrical Engineering and Computer Science, Slovenia}
\affiliation{University of Edinburgh, Usher Institute, United Kingdom}

\author{Roger~Watson \orcidlink{0000-0001-8040-7625}}
\affiliation{Southwest Medical University, China }

\author{Dean~Korošak \orcidlink{0000-0003-3818-1233} }
\thanks{Corresponding author:\\dean.korosak@um.si}
\affiliation{University of Maribor, Faculty of Medicine, Slovenia}
\affiliation{University of Maribor, Faculty of Civil Engineering, Transportation Engineering and Architecture, Maribor, Slovenia}

\date{\today}

\begin{abstract}
The current system of scholarly publishing is often criticized for being slow, expensive, and not transparent. The rise of open access publishing as part of open science tenets, promoting transparency and collaboration, together with calls for research assesment reforms are the results of these criticisms. The emergence of new open access publishers presents a unique opportunity to empirically test how universities and countries respond to shifts in the academic publishing landscape. These new actors challenge traditional publishing models, offering faster review times and broader accessibility, which could influence strategic publishing decisions.

Our findings reveal a clear division in European publishing practices, with countries clustering into two groups distinguished by the ratio of publications in new open access journals with accelerated review times versus legacy journals. This divide underscores a broader shift in academic culture, highlighting new open access publishing venues as a strategic factor influencing national and institutional publishing practices, with significant implications for research accessibility and collaboration across Europe.
\end{abstract}

\maketitle 

\section{Introduction}
One of the mainstays of evaluating the performance of universities is their performance in research, and a major plank of that evaluation is constituted by publication in academic journals. Likewise, the metrics-based evaluation of individual academics follows similar processes shaped by pressures to publish in high-impact journals. Researchers may choose open access venues to increase visibility and compliance with open science mandates, while universities and national science systems might adapt their evaluation criteria, balancing prestige with the growing importance of transparency and public access. This dynamic provides fertile ground for studying how institutions adjust their incentives and how these shifts affect researchers publishing strategies.

Researchers often prioritize journal prestige and citation counts, sometimes at the cost of research quality and broader societal impact. Institutional policies and national funding systems frequently reward publication volume and impact factor, reinforcing this trend. Recent studies show a growing disconnect between researchers' values, which may favor openness and integrity, and institutional incentives focused on metrics. Performance-based funding models and evolving open science practices reveal how current systems reshape research behavior, raising concerns about the sustainability and ethics of research evaluation. This paper examines the impact of these forces on publishing practices and researcher behavior. We have previously discussed the controversies involved in using publication metrics in academic journals to evaluate individual academics~\cite{watson2023assessing} and many of those issues applying to the use of publication in academic journals apply to the evaluation of universities.

Building on the pressures associated with metrics-based evaluations, recent studies suggest that the dominance of journal impact factors and the resulting “publish or perish” culture may further undermine research quality. Bohorquez et al.~\cite{bohorquez2024researchers} found that pressure to publish in prestigious journals can lead researchers to adjust findings to fit publication standards, often at the expense of comprehensive evidence. Researchers under significant pressure may prioritize journal prestige and rapid publication timelines over open-access and transparency considerations, as Johann et al.~\cite{johann_impact_2024} describe, opting for instrumental rather than normative publication strategies. Furthermore, Ross-Hellauer et al.~\cite{ross-hellauer_value_2024} examined the phenomenon of “value dissonance,” where researchers’ commitment to open and responsible research increasingly conflicts with institutional demands for high-impact publications, favoring citation metrics over collaborative and ethical practices. Additionally, Baccini et al.~\cite{baccini2019citation} showed how bibliometric-driven evaluations encourage behaviors like self-citations and strategic citation practices, ultimately fostering a citation-centric approach that may detract from genuine scientific impact. These findings underscore the need to reassess research assessment policies, highlighting the growing gap between institutional metrics and researchers’ values, as well as the importance of aligning evaluation criteria with the principles of open science and research integrity.

The interplay between country-level research reforms and researchers’ publishing choices was highlighted in studies by Cernat~\cite{cernat_unprincipled_2024} and Dagienė et al.~\cite{dagiene_incentivising_2024}, revealing how policy-driven metrics reshape academic behavior. Cernat’s analysis of Romania’s 2016 reforms, which imposed strict publication criteria amidst funding cuts, led to a focus on high-impact journals at the expense of conference proceedings, ultimately reducing overall research productivity. This case exemplifies the misalignment between top-down policy intentions and researchers’ capabilities under constrained resources. Similarly, Dagienė et al.’s study on Lithuania’s performance-based funding system shows how the push for indexed journal publications has spurred strategic publishing behaviors, emphasizing quantity over quality. These cases illustrate the influence of diverse stakeholders—scientific elites, policymakers, universities, and researchers—in shaping research assessment policies, raising concerns about the sustainability and genuine innovation fostered by metrics-driven funding.

Here, we investigate the evolution of academic publishing practices in Europe by analyzing the publishing data on universitiy and country level in the European Union, focusing on the ratio between publications in new, open access journals (MDPI in our case), and those in traditional, legacy journals (here, The Big Five). By examining the distributions of this ratio, we identify two distinct groups of universities and countries with different scholarly publishing cultures. We show how publishing choice correlates with innovation potential and corruption perception at the country level, revealing the broader socio-economic context that shapes academic publishing practices. Our findings reveal significant insights into the current state and evolving trends of scholarly publishing practices among universities and EU countries.

\section{Methods and results}
This study employs open, publicly accessible data from Open Alex~\cite{priem2022openalex} and uses CTWS Open Ranking~\cite{cwts2024leiden} of universities to systematically analyze publishing behaviors across academic institutions. By leveraging these resources, we aim to provide a clear, open data-driven view of trends in scientific publication, highlighting variations in open access practices and publication types across institutions and publishers.

OpenAlex is an open-access platform designed to index scientific publications. It provides access to metadata about scientific papers, journals, authors, and ultimately institutions. The OpenAlex API was utilized for simplified and automatized access to data in the OpenAlex repository. The OpenAlex API is freely available and does not necessitate an API key for utilization, making it fairly easy to use. Due to the rate limits, which are implemented to prevent abuse and ensure fair usage, we had to send requests in 30 second intervals. 

We analyzed the publishing data for all universities that were ranked in CTWS Open Ranking. For each institution, we fetched its data based on ROR ID~\cite{ROR} and publication year. This dataset was then further processed by considering the publication type and calculating the total number of publications. By aggregating ROR ID, year, and journal type, we derived new features such as the number of publications published with MDPI, Taylor\&Francis, Springer Nature, Wiley, Sage, and Elsevier, the number of retracted publications, the number of open access publications, and the number of gold open access publications. These features can be used to examine various indicators of publishing habits and serve as a starting point for comparing institutions publishing culture. 

We selected MDPI as a representative open access publisher because it is successful, relatively new, widely used and offers journals across almost all subjects as a comparison with legacy journals published by more established publishers. All publishers now offer open access publishing, and established publishers, such as Wiley, Elsevier and Springer Nature publish a suite of hybrid journals offering both pay to view and pay to publish alongside a developing suite of open access journals. MDPI, on the other hand, offer only open access journals and compared with the Big Five (Springer Nature, Wiley, Elsevier, Taylor\&Francis, Sage) publishers offer lower article processing charges (APCs) and higher rates of acceptance~\cite{fillon2024should}. 

Here, we define and focus on the ratio, $\rho$, between the number of publications published with MDPI and the Big Five in a given year defined as: 

\begin{equation}
\rho = \frac{N_{\text{MDPI}}}{N_{\text{Big Five}}+N_{\text{MDPI}}},
\end{equation}

where $N_{\text{MDPI}}$ is the number of publications published with MDPI and $N_{\text{Big Five}}$ is the number of publications published with the Big Five publishers.

\begin{figure}
\centering
\includegraphics[width=1.0\linewidth]{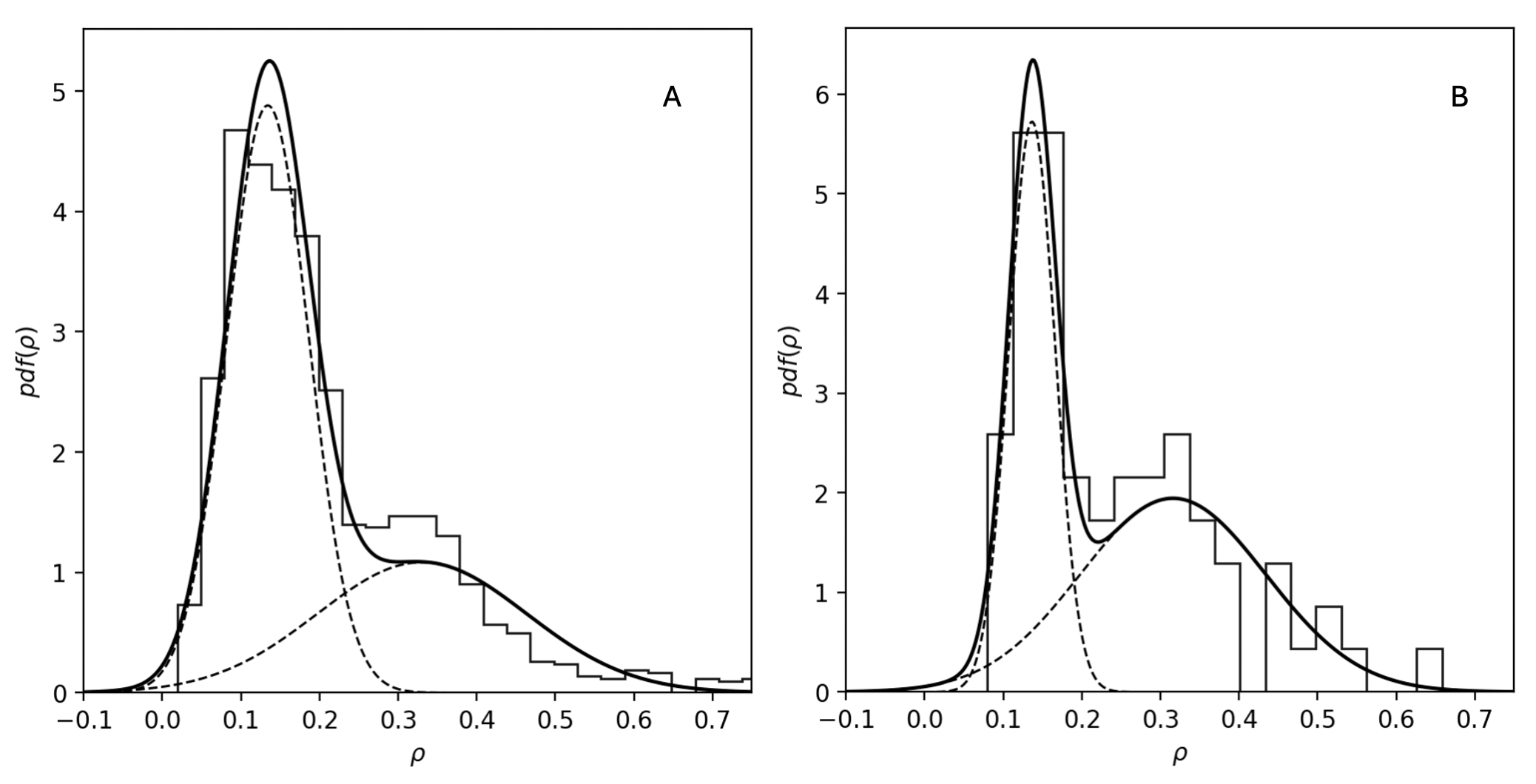}
\caption{\label{fig:fig1} {\bf Distributions of publications ratio at university and country level.} (A) $\rho$ (see eq. 1) distribution for 2022 at the university level, Full line shows two-gaussian mixture fit, individual gaussians are plotted with dashed lines. (B) $\rho$ distributions for 2022 at the country level.
}
\end{figure}

In figure 1 we show the $\rho$ distribution with universities ranked in CWTS Leiden Open Ranking (panel A) for 2022. The distribution is well described by two-gaussian mixture $P(x) = \omega\mathcal{N}(x|\mu_1, \sigma_1) + (1-\omega)\mathcal{N}(x|\mu_2, \sigma_2)$ peaked around mean values $\mu_1, \mu_2$ for $\rho$s with widths $\sigma_1, \sigma_2$. $\omega$ and $1-\omega$ are the weights of each gaussian peak in the mixture distribution. In panel B of Figure 1 we show the $\rho$ distribution across countries with universities ranked in CWTS Leiden Open Ranking. The difference $\mu_2-mu_1$ in gaussian means is increasing with time between 2019 and 2023 as: 0.116, 0.186, 0.219, 0.196, 0.199, at the university level, and as: 0.086, 0.139, 0.174, 0.179, 0.232 at the country level. In both cases the separation between the two groups is increasing over time. 

\begin{figure}
    \centering
    \includegraphics[width=1.0\linewidth]{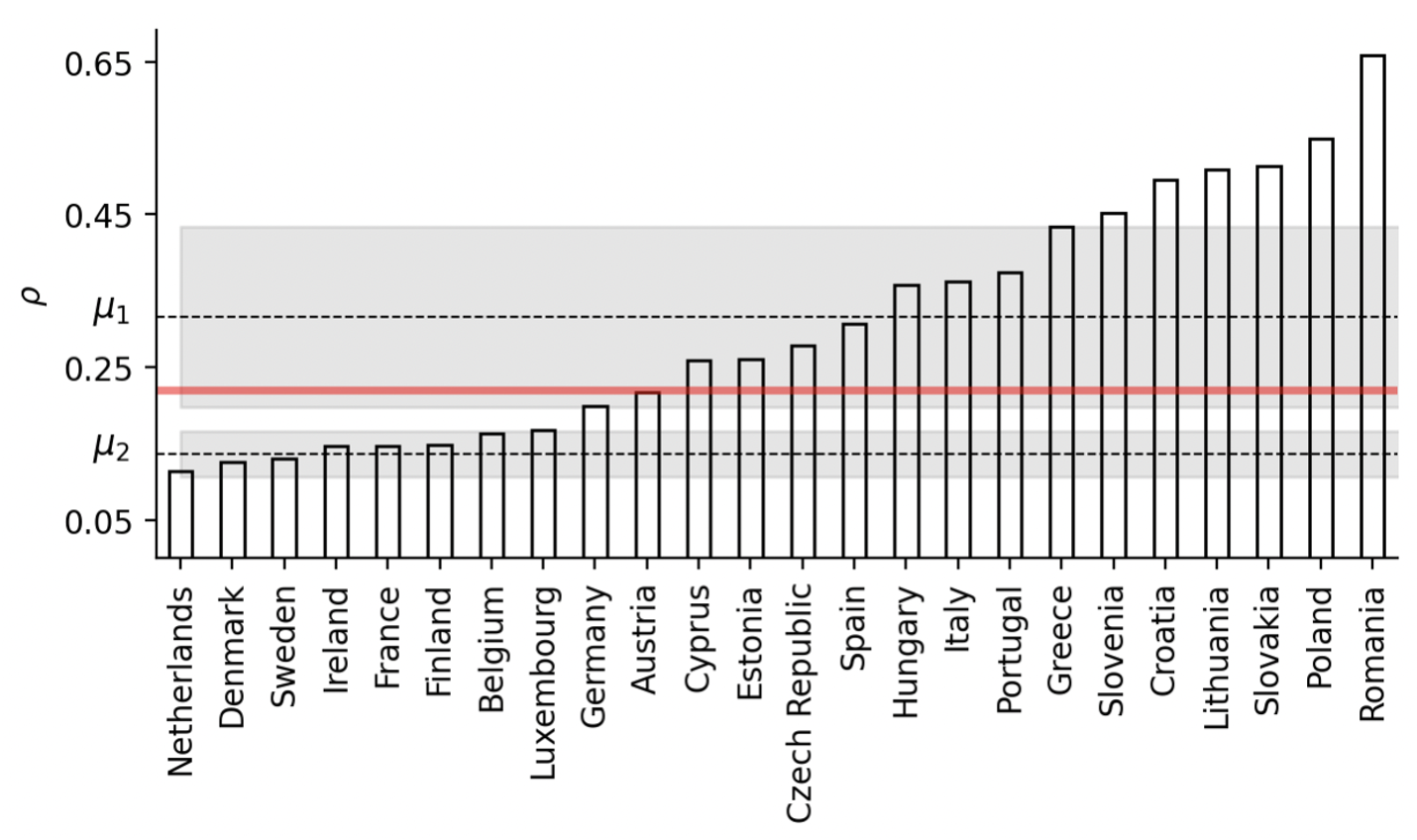}
    \caption{\label{fig:fig2} {\bf Publications ratio at country level.} $\rho$ for 2022 for countries with universities ranked in CWTS Leiden Open Ranking. The thick horizontal line corresponds to the minimum of the two-gaussian mixture fit. The dotted lines are the means $\mu_1$ and $\mu_2$ of the fitted two gaussians and the shaded bands are the corresponding widths $\sigma_1$ and $\sigma_2$ of the distributions.
    }
    \end{figure}

To see which countries comprise each of the two groups, we plotted the $\rho$ for 2022 vs countries with universities ranked in CWTS Leiden Open Ranking in Figure 2. The thick horizontal line corresponds to minimum of the two-gaussian mixture fit and denotes the approximate separation of countries into two groups. The dotted lines are the means $\mu_1$ and $\mu_2$ of the mixed distribution model and the shaded bands are the corresponding standard deviations $\sigma_1$ and $\sigma_2$ of the two peaks. 

The data in Figure 2 indicates a clear separation of countries into two groups: $\rho_{high}$, countries with a higher $\rho$ moslty overlapping with the set of central and south-eastern EU countries (Croatia, Cyprus, Czech Republic, Estonia, Greece, Hungary, Italy, Lithuania, Poland, Portugal, Romania, Slovakia, Slovenia, Spain) and $\rho_{low}$, countries with a lower $\rho$ moslty overlapping with the set of central and north-western EU countries (Austria, Belgium, Denmark, Finland, France, Germany, Ireland, Luxembourg, Netherlands, Sweden). 

To see if the division of countries into $\rho_{high}$ and $\rho_{low}$ groups is statistically significant, we performed a Mann-Whitney U test. The test showed that the mean $\rho$ for $\rho_{high}$ countries is significantly higher than for $\rho_{low}$ countries with $p<0.01$. Figure 3 displays the mean $\rho$ separately for $\rho_{high}$ and $\rho_{low}$ countries. The gap and the increasing trend of the gap over the years in mean $\rho$ between these two country groups is clearly visible. 

\begin{figure}
    \centering
    \includegraphics[width=1.0\linewidth]{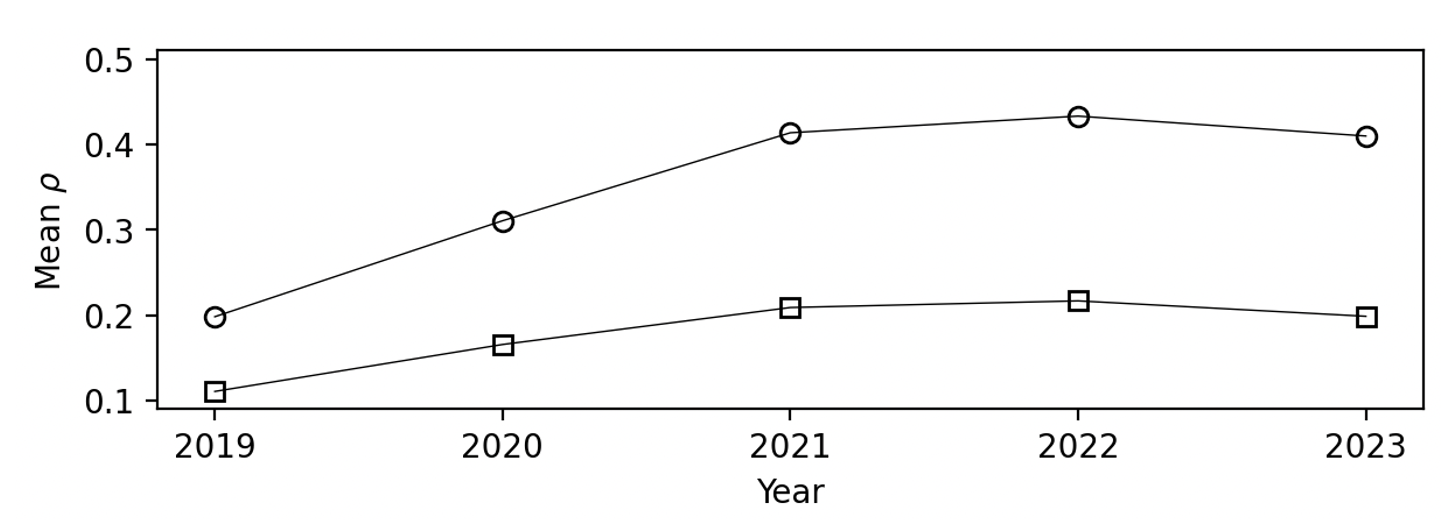}
    \caption{\label{fig:fig3} {\bf Evolution of publications ratio.} Mean $\rho$ vs years for $\rho_{high}$ (open circles) and $\rho_{low}$ (open squares) countries. Mann-Wittney U test shows statistically significant  differences in mean $\rho$ between $\rho_{high}$ and $\rho_{low}$ countries with $p<0.01$.
    }
\end{figure}

\begin{figure}
    \centering
    \includegraphics[width=1.0\linewidth]{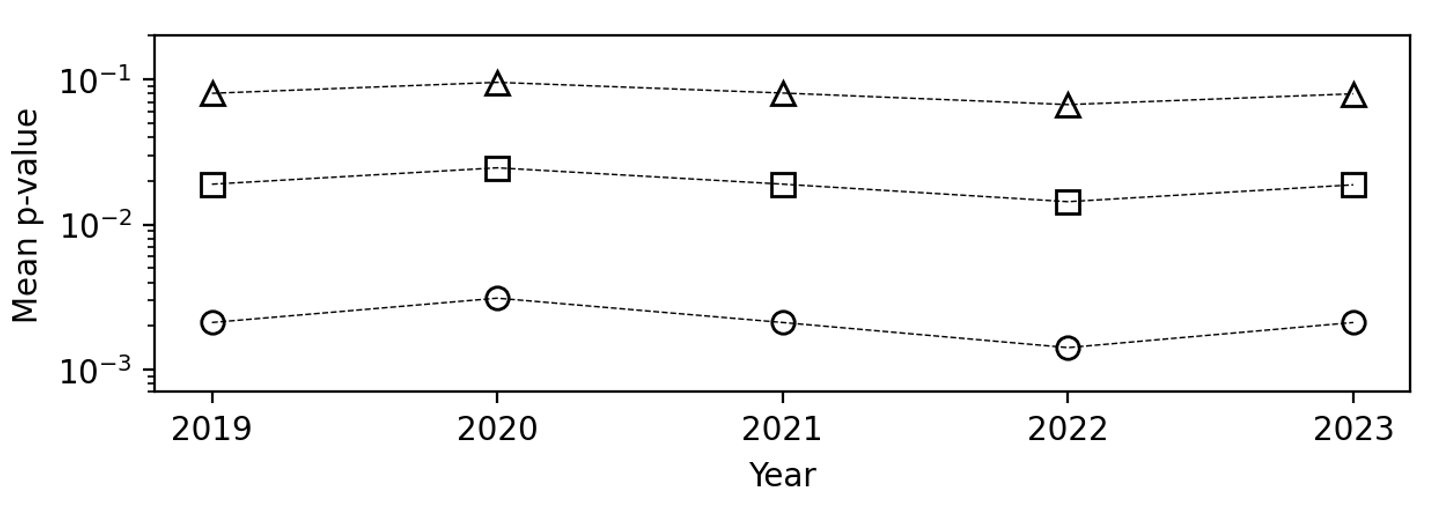}
    \caption{\label{fig:fig4} {\bf Statistical significance of $\rho_{high}$ and $\rho_{low}$ separation.} Mean p-values for Mann-Whitney U test for differences in mean $\rho$ between $\rho_{high}$ and $\rho_{low}$ countries. The p-values are calculated for 10000 random permutations of countries between the two groups in which 0 (open circles), 1 (open squares) or 2 (open triangles) countries are randomly switched between the two groups.
    }
\end{figure}

We also tested the robustness of the separation of countries into $\rho_{high}$ and $\rho_{low}$ groups by randomly switching countries between the two groups. Figure 4 shows the mean p-values for the Mann-Whitney U test for the differences in mean $\rho$ between $\rho_{high}$ and $\rho_{low}$ countries in which 0, 1 or 2 countries are randomly switched between the two groups. The p-values are calculated for 10000 such random permutations of countries between the two groups. The mean p-values vary very little between years for each permutation regime. When only one country is switched between the two groups the division between $\rho_{high}$ and $\rho_{low}$ countries is still statistically significant, however when switching two countries the division is no longer statistically significant showing the robustness of the separation of countries into $\rho_{high}$ and $\rho_{low}$ groups. 

\begin{figure}
    \centering
    \includegraphics[width=1.0\linewidth]{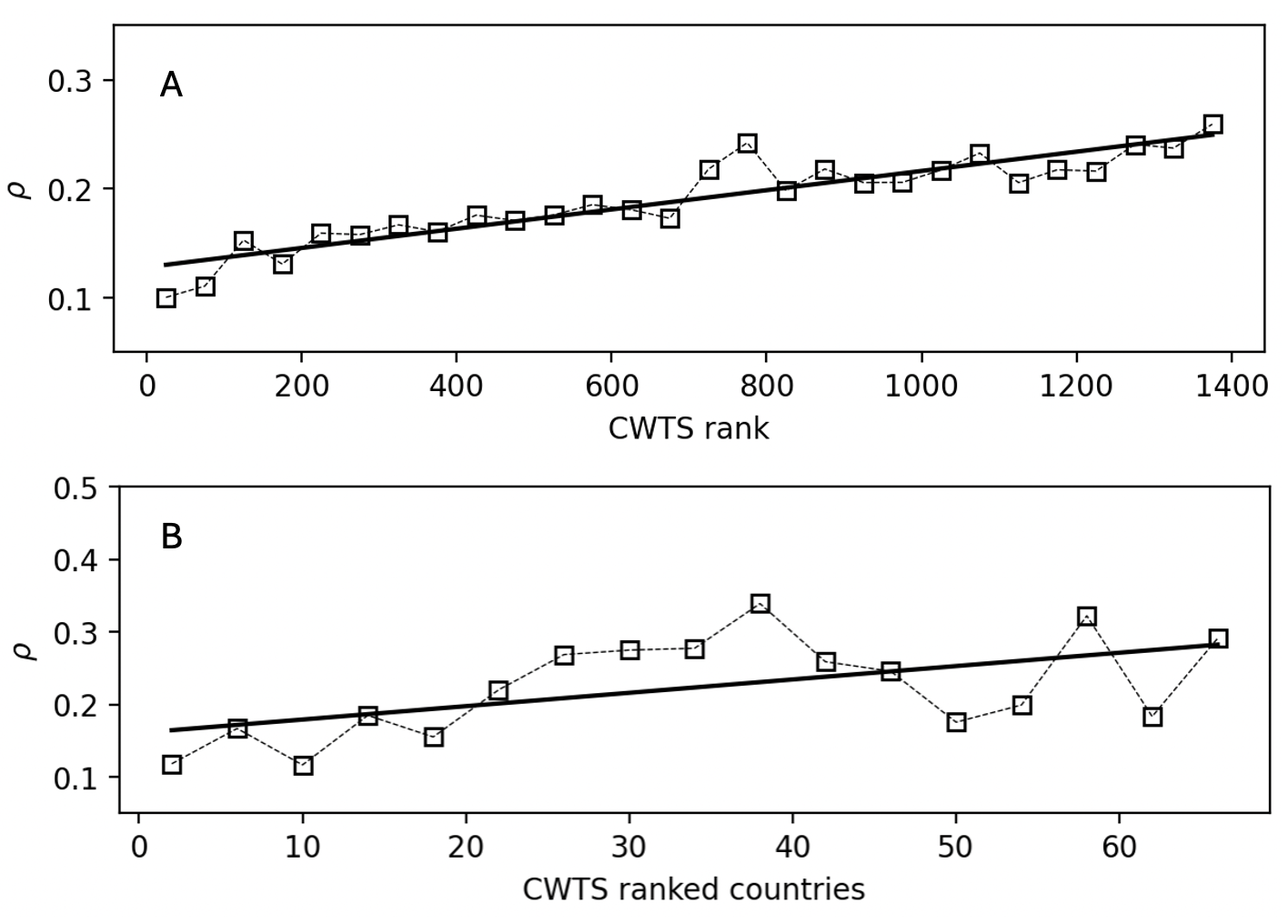}
    \caption{\label{fig:fig5} {\bf Publication ratio and ranking.} Relationship between $\rho$ and the rank of universities (panel A) and countries (panel B) in CWTS Leiden Open Ranking. The individual data points in the figure are the $\rho$s of universities computed as averages in bins 50 ranks wide.  
}
\end{figure}

Research output measured through scholarly publications is one of the key indicators in almost all university rankings so it should not be surprising that the scholarly publishing culture of universities might be closely related to their ranking. We checked this by comparing the $\rho$ of universities with their rank in CWTS Leiden Open Ranking. In figure 5, panel A, we display the relationship between the $\rho$ and the rank of universities included in CWTS Leiden Open Ranking table where the individual data points in the figure are the averages of $\rho$s of universities over intervals of 50 ranks. This relationship can be well described by a simple linear relationship between the rank of the university and its scholarly publishing culture. In panel B of figure 5 we show the relationship between the $\rho$ and the rank of countries of universities included in CWTS Leiden Open Ranking table. Here the relationship is not as clear as in the case of universities, however, there is a trend of higher ranked countries having lower $\rho$.

\begin{figure}
    \centering
    \includegraphics[width=1.0\linewidth]{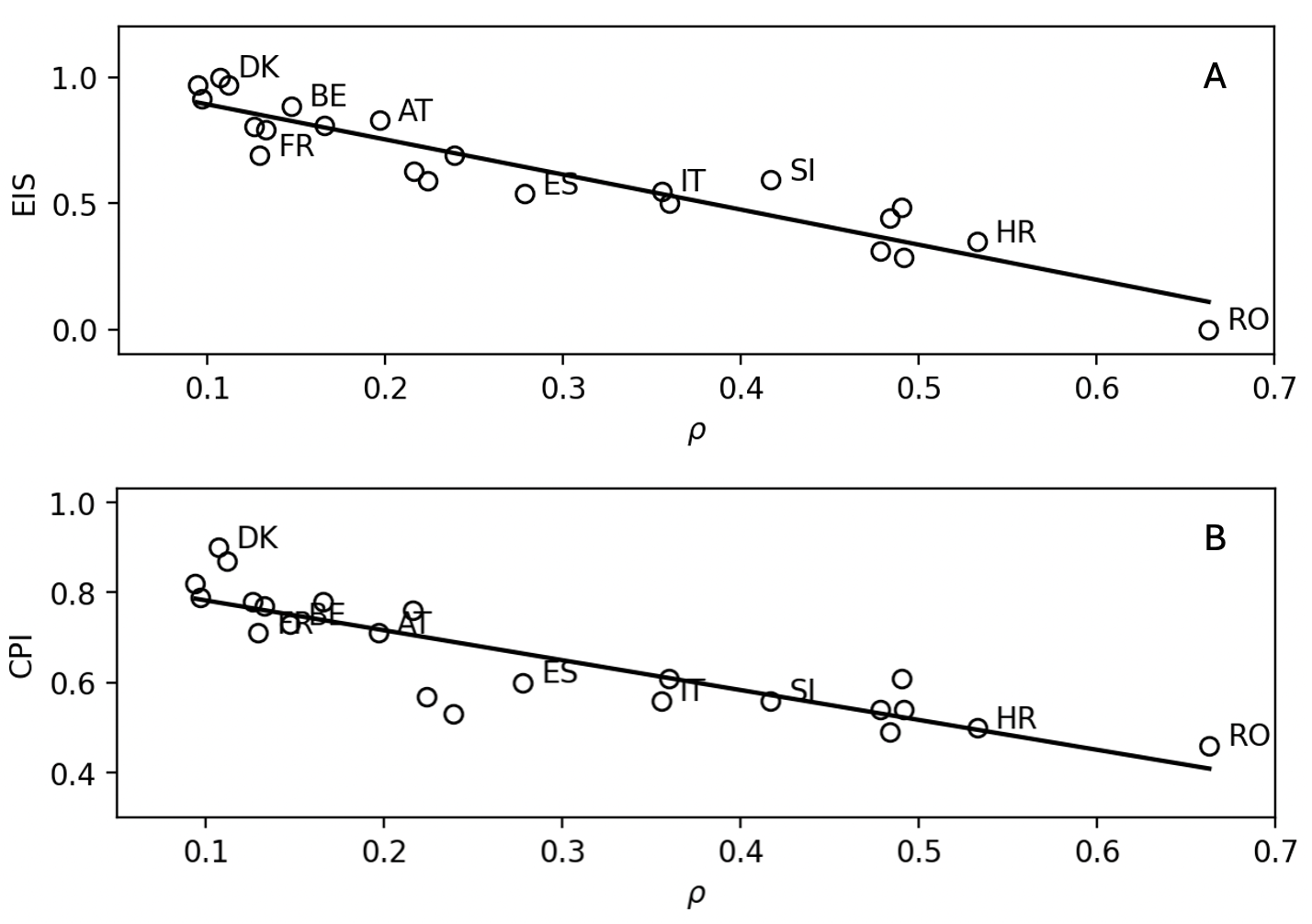}
    \caption{\label{fig:fig6} {\bf Publications ratio relationship with innovations and corruption.} Innovation potential, corruption perception and scholarly publishing culture. (A) Relationship between the normalized European Innovation Scoreboard (EIS) and $\rho$ at a country level. (B) Relationship between the normalized Corruption Perception Index (CPI) and $\rho$ at a country level. For clarity only some of countries are indicated with country codes.
}
\end{figure}

Scholarly publishing culture of countries can also be related to their socio-economic context. Bringing together the data on innovation potential (European Innovation Scoreboard, EIS)~\cite{european_innovation_scoreboard_2023} and corruption perception (Corruption Perception Index, CPI)~\cite{transparency_international_cpi_2022} with the $\rho$ of countries we can see how these factors are related to the scholarly publishing culture of countries. In figure 6 we show the relationship between the (normalized) EIS (panel A), CPI (panel B) and $\rho$ $\rho$ at a country level. For clarity only some of countries are indicated with country codes.
There are clear and strong correlations between $\rho$ and EIS and between $\rho$ and CPI, with correlation coeffients $a=\Corr(\rho, EIS)=-0.93$ and $b=\Corr(\rho, CPI)=-0.86$ respectively, computed from the datasets. Since for EIS higher values indicate better innovation potential and for CPI higher values indicate lower corruption perception, the strong correlations between $\rho$ and EIS and between $\rho$ and CPI indicate that countries with higher innovation potential and lower corruption perception are associated with lower $\rho$. 

The strong correlations between $\rho$ and EIS and between $\rho$ and CPI serve also as a validation of choosing the $\rho$ as a measure of scholarly publishing culture in the sense that due to correlation transitivity in such case one should also observe a strong positive correlation between EIS and CPI. Indeed, the empirical correlation between EIS and CPI is $\Corr(EIS, CPI)=0.87$, while the theoretical lower bound for the correlation coefficient between EIS and CPI is $ab-\sqrt{1-a^2}\sqrt{1-b^2}=0.61$.

We showed that the $\rho$ distribution in case of universities and countries is well described by two-gaussian mixed distribution which is a weighted sum of two gaussian distributions. Suppose that means and variances evolve with time much slower than the values of the weight $\omega\in [0, 1]$.  Therefore, the shape of $P(x)$ will be mostly determined by the time evolution of $\omega$. The emergence of two cultures in scholarly publishing when $\omega > 0$ and when the second peak with $\mu_2 >\mu_1$ occurs.  

Let us consider a simple game theoretical model to explain the possible emergence of two cultures in scholarly publishing.
Strategic publishing choices by researchers are mostly driven by current "publish or perish" culture pervailing in academia. Prisoner’s dilemma type of a game was used~\cite{publish-and-flourish_2019,erren_analyzing_2016} to illustrate the dominant strategy to defect by hyping or exaggerating research findings to secure publication in high ranking journals over honest reporting, a behavior that undermines scientific integrity and collaboration. In this strategic game two individuals, acting in their self-interest, both choose to defect, leading to a worse outcome for both than if they had cooperated—resulting in a stable but suboptimal equilibrium. Researchers face intense pressure to maximize publication output coupled with prestige, often tied to limited slots in high-impact journals, exacerbates this competitive dynamic.

New open-access publishers, with fast review processes and unlimited publication space, offer a potential escape from this dilemma. By emphasizing transparency, data sharing, and even open peer review, these platforms align more closely with open science principles, prioritizing accessibility and reproducibility over exclusivity. However, without the same level of prestige as traditional journals, these open-access venues often lack the same career-advancing power. Researchers, therefore, remain in a bind, pulled between the values of open science and the prestige requirements of their institutions.

Let $\omega$ be the fraction of researchers with the strategy to cooperate by choosing open-access publication. Consequently, $1 - \omega$ is the fraction adhering to legacy publication, tied to the original defection equilibrium in the Prisoner’s Dilemma. Each researcher’s decision (defect or cooperate) contributes to an aggregate behavior that shapes the observed distribution of $\rho$. To capture the population aspect, we consider the game as a decision choice by a single researcher and an “average researcher” who embodies current publication culture, incentives, policies and which reflects the payoff structure that an individual researcher faces when deciding whether to defect (publish in legacy journals) or cooperate (publish in open access). 

Here we apply the replicator equation~\cite{taylor1978evolutionary} to describe how cooperation (open access publishing) and defection (legacy publishing) strategies evolve within a population based on their relative payoffs, i.e. how the fraction of cooperators, denoted by $\omega$, shifts over time, shaped by the academic environment, including policies and pressures that influence publishing choices. Let's use a very simple payoff difference between cooperation and defection $\pi_{\text{cooperate}} - \pi_{\text{defect}} = \lambda - k \omega$, where external factors like policy support for open access are represented by a parameter $\lambda$, and $k > 0$ is a constant capturing the effect of diminishing returns as $\omega$ increases since as more researchers publish in open access, the payoff advantage might decrease due to saturation or competition.

Using this payoff difference in a simple replicator dynamic yields the equation for the rate of change in the fraction of researchers choosing cooperation: $\frac{d\omega}{dt} = \omega (1 - \omega)(\lambda - k \omega)$. The equilibria (fixed points) are the three solutions of the steady state $\frac{d\omega}{dt} = 0$. These are: $\omega = 0$, everyone defects (publishes in legacy journals), $\omega = 1$, everyone cooperates (publishes in open access), and $\omega = \frac{\lambda}{k}$, a mixed equilibrium where a fraction of researchers cooperates.

The behavior of the system depends on the value of $\lambda$. If $\lambda < k$, the term $\frac{\lambda}{k}$ lies between 0 and 1, resulting in three equilibria: two stable ones ($\omega = 0$ and $\omega = 1$) and one unstable equilibrium at $\omega = \frac{\lambda}{k}$. If $\lambda > k$, the only stable equilibrium is $\omega = 1$, meaning that cooperation becomes the dominant strategy for the entire population as the payoff difference strongly favors open access.

As $\lambda$ crosses a critical threshold (equal to k), the system shifts from having a single stable equilibrium (everyone defecting) to multiple equilibria (defection and cooperation coexisting). For low $\lambda$ the population favors legacy publishing due to low support for open access, $\omega$ remains close to 0, and the distribution of $\rho$ shows a single peak. High $\lambda$ values models strong policy support that make open-access publishing attractive, pushing the system toward $\omega = 1$, where everyone cooperates, resulting in a single peak at a high $\rho$. At intermediate values of $\lambda$ the system allows coexistence, where part of the population defects and the other part cooperates, corresponding to a mixed distribution with two peaks in $\rho$. This transition is analogous to a phase transition in physical systems, where small changes in external conditions lead to a qualitative shift in the system's behavior. Here, it represents a shift in the academic community from predominantly legacy publishing to a coexistence of open-access and legacy publishing as viable choices.

This bifurcation is observed in empirical data as a mixed distribution with two peaks, corresponding to the two strategies. When $\lambda$ is low, the system favors legacy publishing, with most researchers defecting. As $\lambda$ increases and approaches the threshold, a second peak emerges as cooperation becomes more attractive, leading to a population split between those who favor open access and those who remain with legacy journals. This mixed distribution reflects the aggregate outcome of individual decisions made within the evolving landscape of academic publishing, driven by socio-economic incentives and game-theoretic principles encapsulated in the replicator dynamics.

\section{Discussion}
Other have shown before us that there are differences in publication patterns across european countries~\cite{kulczycki_publication_2018,sasvari2023current,csomos_understanding_2023}. 
Our findings add to and extend these results, and reveal significant new insights into the current state and evolving trends of scholarly publishing practices among universities and EU countries. Using open research information of scholarly publications and university ranking we showed that there is a clear bifurcation in publishing cultures in Europe. 

The separation, evident at both the university and country levels, indicates a growing divergence in publication strategies, influenced by factors such as policy changes and institutional rankings. The increasing gap between $\rho_{high}$ and $\rho_{low}$ countries underscores the impact of regional dynamics on academic publishing. Additionally, the correlation between publishing practices and factors like innovation potential and corruption perception suggests that broader socio-economic contexts play a role in shaping research outputs. 

The tendency of researchers from $\rho_{high}$ countries and lower-ranked institutions to publish more in open-access MDPI journals, which provide faster turnaround and are often perceived as less stringent, may be explained in several ways. As publishing behavior is generally the result of a reasoned process, our findings can be interpreted through the lens of the Theory of Planned Behavior~\cite{ajzen1991}, which recognizes the critical role of individuals ability (i.e., perceived behavioral control) and motivation (i.e., intentions, which are heavily influenced by subjective norms and attitudes). While this theory has not yet been studied in the context of explaining researchers’ decision to publish in MDPI over more traditional journals, Moksness and colleagues~\cite{moksness2020} have used it to explain intentions to publish in open-access journals.

First, in relation to perceived behavioral control, researchers from $\rho_{high}$ and lower-ranked countries may perceive important internal and external barriers to publishing in traditional, legacy journals, such as resource constraints (including financial aspects), fewer international collaboration opportunities, language barriers, and less prestigious institutional affiliations. In fact, recent research suggests that some of these barriers objectively exist; for example, a recent study by Sverdlichenko and colleagues~\cite{sverdlichenko2022} revealed that journal editors may be influenced by author institutional affiliations when deciding whether a manuscript should be sent out for peer review. Second, regarding subjective norms, academics working in $\rho_{high}$ and lower-ranked countries may feel a stronger social pressure to publish frequently, potentially driven by institutional policies that still reward the quantity (instead of quality) of publications. This may lead to a snowball effect, with early adopters of such practices imposing pressure on others to do the same to remain competitive in their academic environment. Over time, publishing in MDPI over legacy journals can become completely normalized or even desirable in certain academic environments. The important role of social norms, as opposed to solely top-down regulations, in determining researchers' publication choices has been previously discussed by other authors (e.g., Migheli and Ramello~\cite{migheli2013}). Lastly, several aspects may contribute to more favorable attitudes towards MDPI in $\rho_{high}$ and lower-ranked countries. For example, they may believe that publishing in MDPI will help them achieve their academic goals, such as career advancement, quicker, or increase their visibility (due to the open-access nature of these journals).

Our results connected publishing patterns with socio-economic indicators. We found that researchers from countries with higher corruption perception are more likely to publish in MDPI over the Big Five journals. While this relationship has not yet been investigated in other countries, some parallels can be drawn from the broader literature on the characteristics of individuals who live in countries perceived as highly corrupt. For example, it is well-known that perceptions of corruption are associated with lower institutional trust~\cite{hakhverdian2012}, lower meritocratic ideology~\cite{tan2017}, and adaptive behaviors, whereby societal levels of corruption result in norm and rule violations on the individual level~\cite{kobis2018}. In the specific context of scholarly publishing, researchers may distrust the fairness of the peer-review processes and traditional publishing practices and merits in general. Moreover, environments with high corruption may sometimes be characterized by uncertain career advancement procedures and less transparent and equitable funding, motivating researchers to publish quickly to bolster their CVs and improve their chances in these ambiguous circumstances. 

The observed relationship between EIS, CPI, and $\rho$ highlights how academic publishing choices can mirror broader socio-economic and governance factors, with $\rho$ acting as a significant indicator of these underlying dynamics. Countries with higher innovation (EIS) and lower corruption (CPI) may be more entrenched in established academic practices, favoring legacy journals, while countries with lower scores may rely more heavily on newer, open access journals, possibly due to less stringent academic systems, fewer resources, or a push for faster academic output. The choice to publish in newer, fast-review open access journals versus legacy journals might also be tied to institutional strength and governance quality. 

Our findings raise concerns about the potential inequalities in scholarly publishing across Europe. The preference for MDPI publications in $\rho_{high}$ countries and lower-ranked universities may be influenced by factors such as publication speed and perceived ease of acceptance but could also indicate various challenges faced by researchers in these environments, such as fewer resources available for research and systemic issues that push researchers towards certain publication outlets. In line with this, governments and institutions in $\rho_{high}$ countries should continue enhancing their support for research activities (e.g., transparent funding procedures, budgets for article processing charges) and empowering academics with better infrastructure, mentoring, and collaboration options. Moreover, extensive efforts should be dedicated to promoting transparent and fair evaluation metrics that go beyond the mere quantity of publications.

\begin{center}
--\,--\,--\,--\,--
\end{center}
\vspace{1mm}
\noindent\textbf{Acknowledgements.} DK received financial support from the Slovenian Research Agency (the research core funding program P3-0396 and the research project no.J7-3156). 

\noindent\textbf{Code and data availability.} The data and the code used in this work is available at \url{https://github.com/deankorosak/two-cultures/}.

\noindent\textbf{Author contributions.} All authors contributed substantially to all aspects of the study.

\noindent\textbf{Conflict of interest.} The authors declare no conflict of interest, financial or otherwise.

\bibliography{refs}{}

\begin{thebibliography}{26}%
\makeatletter
\providecommand \@ifxundefined [1]{%
 \@ifx{#1\undefined}
}%
\providecommand \@ifnum [1]{%
 \ifnum #1\expandafter \@firstoftwo
 \else \expandafter \@secondoftwo
 \fi
}%
\providecommand \@ifx [1]{%
 \ifx #1\expandafter \@firstoftwo
 \else \expandafter \@secondoftwo
 \fi
}%
\providecommand \natexlab [1]{#1}%
\providecommand \enquote  [1]{``#1''}%
\providecommand \bibnamefont  [1]{#1}%
\providecommand \bibfnamefont [1]{#1}%
\providecommand \citenamefont [1]{#1}%
\providecommand \href@noop [0]{\@secondoftwo}%
\providecommand \href [0]{\begingroup \@sanitize@url \@href}%
\providecommand \@href[1]{\@@startlink{#1}\@@href}%
\providecommand \@@href[1]{\endgroup#1\@@endlink}%
\providecommand \@sanitize@url [0]{\catcode `\\12\catcode `\$12\catcode `\&12\catcode `\#12\catcode `\^12\catcode `\_12\catcode `\%12\relax}%
\providecommand \@@startlink[1]{}%
\providecommand \@@endlink[0]{}%
\providecommand \url  [0]{\begingroup\@sanitize@url \@url }%
\providecommand \@url [1]{\endgroup\@href {#1}{\urlprefix }}%
\providecommand \urlprefix  [0]{URL }%
\providecommand \Eprint [0]{\href }%
\providecommand \doibase [0]{http://dx.doi.org/}%
\providecommand \selectlanguage [0]{\@gobble}%
\providecommand \bibinfo  [0]{\@secondoftwo}%
\providecommand \bibfield  [0]{\@secondoftwo}%
\providecommand \translation [1]{[#1]}%
\providecommand \BibitemOpen [0]{}%
\providecommand \bibitemStop [0]{}%
\providecommand \bibitemNoStop [0]{.\EOS\space}%
\providecommand \EOS [0]{\spacefactor3000\relax}%
\providecommand \BibitemShut  [1]{\csname bibitem#1\endcsname}%
\let\auto@bib@innerbib\@empty
\bibitem [{\citenamefont {Watson}\ \emph {et~al.}(2023)\citenamefont {Watson}, \citenamefont {Korošak},\ and\ \citenamefont {Štiglic}}]{watson2023assessing}%
  \BibitemOpen
  \bibfield  {author} {\bibinfo {author} {\bibfnamefont {R.}~\bibnamefont {Watson}}, \bibinfo {author} {\bibfnamefont {D.}~\bibnamefont {Korošak}}, \ and\ \bibinfo {author} {\bibfnamefont {G.}~\bibnamefont {Štiglic}},\ }\href@noop {} {\enquote {\bibinfo {title} {Assessing individual research performance},}\ }\bibinfo {howpublished} {\url{https://www.hepi.ac.uk/2023/05/02/assessing-individual-research-performance/}} (\bibinfo {year} {2023}),\ \bibinfo {note} {{HEPI Blog}}\BibitemShut {NoStop}%
\bibitem [{\citenamefont {Bohorquez}\ \emph {et~al.}(2024)\citenamefont {Bohorquez}, \citenamefont {Weerasuriya}, \citenamefont {Brain}, \citenamefont {Senanayake}, \citenamefont {Kularatna},\ and\ \citenamefont {Barnett}}]{bohorquez2024researchers}%
  \BibitemOpen
  \bibfield  {author} {\bibinfo {author} {\bibfnamefont {N.~G.}\ \bibnamefont {Bohorquez}}, \bibinfo {author} {\bibfnamefont {S.}~\bibnamefont {Weerasuriya}}, \bibinfo {author} {\bibfnamefont {D.}~\bibnamefont {Brain}}, \bibinfo {author} {\bibfnamefont {S.}~\bibnamefont {Senanayake}}, \bibinfo {author} {\bibfnamefont {S.}~\bibnamefont {Kularatna}}, \ and\ \bibinfo {author} {\bibfnamefont {A.}~\bibnamefont {Barnett}},\ }\href@noop {} {\bibfield  {journal} {\bibinfo  {journal} {Clinical Science}\ }\textbf {\bibinfo {volume} {297}},\ \bibinfo {pages} {57} (\bibinfo {year} {2024})}\BibitemShut {NoStop}%
\bibitem [{\citenamefont {Johann}\ \emph {et~al.}(2024)\citenamefont {Johann}, \citenamefont {Neufeld}, \citenamefont {Thomas}, \citenamefont {Rathmann},\ and\ \citenamefont {Rauhut}}]{johann_impact_2024}%
  \BibitemOpen
  \bibfield  {author} {\bibinfo {author} {\bibfnamefont {D.}~\bibnamefont {Johann}}, \bibinfo {author} {\bibfnamefont {J.}~\bibnamefont {Neufeld}}, \bibinfo {author} {\bibfnamefont {K.}~\bibnamefont {Thomas}}, \bibinfo {author} {\bibfnamefont {J.}~\bibnamefont {Rathmann}}, \ and\ \bibinfo {author} {\bibfnamefont {H.}~\bibnamefont {Rauhut}},\ }\href {\doibase 10.1093/reseval/rvae011} {\bibfield  {journal} {\bibinfo  {journal} {Research Evaluation}\ ,\ \bibinfo {pages} {rvae011}} (\bibinfo {year} {2024})}\BibitemShut {NoStop}%
\bibitem [{\citenamefont {Ross-Hellauer}\ \emph {et~al.}(2024)\citenamefont {Ross-Hellauer}, \citenamefont {Klebel}, \citenamefont {Knoth},\ and\ \citenamefont {Pontika}}]{ross-hellauer_value_2024}%
  \BibitemOpen
  \bibfield  {author} {\bibinfo {author} {\bibfnamefont {T.}~\bibnamefont {Ross-Hellauer}}, \bibinfo {author} {\bibfnamefont {T.}~\bibnamefont {Klebel}}, \bibinfo {author} {\bibfnamefont {P.}~\bibnamefont {Knoth}}, \ and\ \bibinfo {author} {\bibfnamefont {N.}~\bibnamefont {Pontika}},\ }\href {\doibase 10.1093/scipol/scad073} {\bibfield  {journal} {\bibinfo  {journal} {Science and Public Policy}\ }\textbf {\bibinfo {volume} {51}},\ \bibinfo {pages} {337} (\bibinfo {year} {2024})}\BibitemShut {NoStop}%
\bibitem [{\citenamefont {Baccini}\ \emph {et~al.}(2019)\citenamefont {Baccini}, \citenamefont {De~Nicolao},\ and\ \citenamefont {Petrovich}}]{baccini2019citation}%
  \BibitemOpen
  \bibfield  {author} {\bibinfo {author} {\bibfnamefont {A.}~\bibnamefont {Baccini}}, \bibinfo {author} {\bibfnamefont {G.}~\bibnamefont {De~Nicolao}}, \ and\ \bibinfo {author} {\bibfnamefont {E.}~\bibnamefont {Petrovich}},\ }\href@noop {} {\bibfield  {journal} {\bibinfo  {journal} {PLoS One}\ }\textbf {\bibinfo {volume} {14}},\ \bibinfo {pages} {e0221212} (\bibinfo {year} {2019})}\BibitemShut {NoStop}%
\bibitem [{\citenamefont {Cernat}(2024)}]{cernat_unprincipled_2024}%
  \BibitemOpen
  \bibfield  {author} {\bibinfo {author} {\bibfnamefont {V.}~\bibnamefont {Cernat}},\ }\href {\doibase 10.1007/s11192-024-05118-9} {\bibfield  {journal} {\bibinfo  {journal} {Scientometrics}\ }\textbf {\bibinfo {volume} {129}},\ \bibinfo {pages} {5557} (\bibinfo {year} {2024})}\BibitemShut {NoStop}%
\bibitem [{\citenamefont {Dagiene}\ \emph {et~al.}(2024)\citenamefont {Dagiene}, \citenamefont {Larivière}, \citenamefont {Dix},\ and\ \citenamefont {Waltman}}]{dagiene_incentivising_2024}%
  \BibitemOpen
  \bibfield  {author} {\bibinfo {author} {\bibfnamefont {E.}~\bibnamefont {Dagiene}}, \bibinfo {author} {\bibfnamefont {V.}~\bibnamefont {Larivière}}, \bibinfo {author} {\bibfnamefont {G.}~\bibnamefont {Dix}}, \ and\ \bibinfo {author} {\bibfnamefont {L.}~\bibnamefont {Waltman}},\ }\href {\doibase 10.31235/osf.io/9yq38} {\  (\bibinfo {year} {2024}),\ 10.31235/osf.io/9yq38}\BibitemShut {NoStop}%
\bibitem [{\citenamefont {Priem}\ \emph {et~al.}(2022)\citenamefont {Priem}, \citenamefont {Piwowar},\ and\ \citenamefont {Orr}}]{priem2022openalex}%
  \BibitemOpen
  \bibfield  {author} {\bibinfo {author} {\bibfnamefont {J.}~\bibnamefont {Priem}}, \bibinfo {author} {\bibfnamefont {H.}~\bibnamefont {Piwowar}}, \ and\ \bibinfo {author} {\bibfnamefont {R.}~\bibnamefont {Orr}},\ }\href@noop {} {\bibfield  {journal} {\bibinfo  {journal} {arXiv preprint arXiv:2205.01833}\ } (\bibinfo {year} {2022})}\BibitemShut {NoStop}%
\bibitem [{\citenamefont {{CWTS, Centre for Science and Technology Studies}}(2024)}]{cwts2024leiden}%
  \BibitemOpen
  \bibfield  {author} {\bibinfo {author} {\bibnamefont {{CWTS, Centre for Science and Technology Studies}}},\ }\href@noop {} {\enquote {\bibinfo {title} {{CWTS Leiden Ranking Open Edition}},}\ }\bibinfo {howpublished} {\url{https://open.leidenranking.com/}} (\bibinfo {year} {2024}),\ \bibinfo {note} {accessed: 2024-06-11}\BibitemShut {NoStop}%
\bibitem [{\citenamefont {{Research Organization Registry}}()}]{ROR}%
  \BibitemOpen
  \bibfield  {author} {\bibinfo {author} {\bibnamefont {{Research Organization Registry}}},\ }\href@noop {} {\enquote {\bibinfo {title} {{Research Organization Registry (ROR)}},}\ }\bibinfo {howpublished} {\url{https://ror.org/}},\ \bibinfo {note} {accessed: 2024-06-11}\BibitemShut {NoStop}%
\bibitem [{\citenamefont {Fillon}\ \emph {et~al.}(2024)\citenamefont {Fillon}, \citenamefont {Maniadis}, \citenamefont {Mendez},\ and\ \citenamefont {Sanchez-Nuneez}}]{fillon2024should}%
  \BibitemOpen
  \bibfield  {author} {\bibinfo {author} {\bibfnamefont {A.}~\bibnamefont {Fillon}}, \bibinfo {author} {\bibfnamefont {Z.}~\bibnamefont {Maniadis}}, \bibinfo {author} {\bibfnamefont {E.}~\bibnamefont {Mendez}}, \ and\ \bibinfo {author} {\bibfnamefont {P.}~\bibnamefont {Sanchez-Nuneez}},\ }\href {\doibase 10.12688/openreseurope.17694.1} {\bibfield  {journal} {\bibinfo  {journal} {Open Res Europe}\ }\textbf {\bibinfo {volume} {4:127}} (\bibinfo {year} {2024}),\ 10.12688/openreseurope.17694.1}\BibitemShut {NoStop}%
\bibitem [{\citenamefont {{European Commission}}(2023)}]{european_innovation_scoreboard_2023}%
  \BibitemOpen
  \bibfield  {author} {\bibinfo {author} {\bibnamefont {{European Commission}}},\ }\href@noop {} {\enquote {\bibinfo {title} {{European Innovation Scoreboard 2023}},}\ }\bibinfo {howpublished} {\url{https://research-and-innovation.ec.europa.eu/statistics/performance-indicators/european-innovation-scoreboard_en}} (\bibinfo {year} {2023}),\ \bibinfo {note} {accessed: 2024-10-02}\BibitemShut {NoStop}%
\bibitem [{\citenamefont {{Transparency International}}(2022)}]{transparency_international_cpi_2022}%
  \BibitemOpen
  \bibfield  {author} {\bibinfo {author} {\bibnamefont {{Transparency International}}},\ }\href@noop {} {\enquote {\bibinfo {title} {{Corruption Perceptions Index 2022}},}\ }\bibinfo {howpublished} {\url{https://www.transparency.org/en/cpi/2022}} (\bibinfo {year} {2022}),\ \bibinfo {note} {accessed: 2024-10-02}\BibitemShut {NoStop}%
\bibitem [{\citenamefont {Stojmenova~Duh}\ \emph {et~al.}(2019)\citenamefont {Stojmenova~Duh}, \citenamefont {Duh}, \citenamefont {Droftina}, \citenamefont {Kos}, \citenamefont {Duh}, \citenamefont {Simonič~Korošak},\ and\ \citenamefont {Korošak}}]{publish-and-flourish_2019}%
  \BibitemOpen
  \bibfield  {author} {\bibinfo {author} {\bibfnamefont {E.}~\bibnamefont {Stojmenova~Duh}}, \bibinfo {author} {\bibfnamefont {A.}~\bibnamefont {Duh}}, \bibinfo {author} {\bibfnamefont {U.}~\bibnamefont {Droftina}}, \bibinfo {author} {\bibfnamefont {T.}~\bibnamefont {Kos}}, \bibinfo {author} {\bibfnamefont {U.}~\bibnamefont {Duh}}, \bibinfo {author} {\bibfnamefont {T.}~\bibnamefont {Simonič~Korošak}}, \ and\ \bibinfo {author} {\bibfnamefont {D.}~\bibnamefont {Korošak}},\ }\href {\doibase 10.3390/publications7020033} {\bibfield  {journal} {\bibinfo  {journal} {Publications}\ }\textbf {\bibinfo {volume} {7}},\ \bibinfo {pages} {33} (\bibinfo {year} {2019})}\BibitemShut {NoStop}%
\bibitem [{\citenamefont {Erren}\ \emph {et~al.}(2016)\citenamefont {Erren}, \citenamefont {Shaw},\ and\ \citenamefont {Morfeld}}]{erren_analyzing_2016}%
  \BibitemOpen
  \bibfield  {author} {\bibinfo {author} {\bibfnamefont {T.~C.}\ \bibnamefont {Erren}}, \bibinfo {author} {\bibfnamefont {D.~M.}\ \bibnamefont {Shaw}}, \ and\ \bibinfo {author} {\bibfnamefont {P.}~\bibnamefont {Morfeld}},\ }\href {\doibase 10.1007/s11948-015-9701-x} {\bibfield  {journal} {\bibinfo  {journal} {Sci Eng Ethics}\ }\textbf {\bibinfo {volume} {22}},\ \bibinfo {pages} {1431} (\bibinfo {year} {2016})}\BibitemShut {NoStop}%
\bibitem [{\citenamefont {Taylor}\ and\ \citenamefont {Jonker}(1978)}]{taylor1978evolutionary}%
  \BibitemOpen
  \bibfield  {author} {\bibinfo {author} {\bibfnamefont {P.~D.}\ \bibnamefont {Taylor}}\ and\ \bibinfo {author} {\bibfnamefont {L.~B.}\ \bibnamefont {Jonker}},\ }\href@noop {} {\bibfield  {journal} {\bibinfo  {journal} {Mathematical biosciences}\ }\textbf {\bibinfo {volume} {40}},\ \bibinfo {pages} {145} (\bibinfo {year} {1978})}\BibitemShut {NoStop}%
\bibitem [{\citenamefont {Kulczycki}\ \emph {et~al.}(2018)\citenamefont {Kulczycki}, \citenamefont {Engels}, \citenamefont {Pölönen}, \citenamefont {Bruun}, \citenamefont {Dušková}, \citenamefont {Guns}, \citenamefont {Nowotniak}, \citenamefont {Petr}, \citenamefont {Sivertsen}, \citenamefont {Istenič~Starčič},\ and\ \citenamefont {Zuccala}}]{kulczycki_publication_2018}%
  \BibitemOpen
  \bibfield  {author} {\bibinfo {author} {\bibfnamefont {E.}~\bibnamefont {Kulczycki}}, \bibinfo {author} {\bibfnamefont {T.~C.~E.}\ \bibnamefont {Engels}}, \bibinfo {author} {\bibfnamefont {J.}~\bibnamefont {Pölönen}}, \bibinfo {author} {\bibfnamefont {K.}~\bibnamefont {Bruun}}, \bibinfo {author} {\bibfnamefont {M.}~\bibnamefont {Dušková}}, \bibinfo {author} {\bibfnamefont {R.}~\bibnamefont {Guns}}, \bibinfo {author} {\bibfnamefont {R.}~\bibnamefont {Nowotniak}}, \bibinfo {author} {\bibfnamefont {M.}~\bibnamefont {Petr}}, \bibinfo {author} {\bibfnamefont {G.}~\bibnamefont {Sivertsen}}, \bibinfo {author} {\bibfnamefont {A.}~\bibnamefont {Istenič~Starčič}}, \ and\ \bibinfo {author} {\bibfnamefont {A.}~\bibnamefont {Zuccala}},\ }\href {\doibase 10.1007/s11192-018-2711-0} {\bibfield  {journal} {\bibinfo  {journal} {Scientometrics}\ }\textbf {\bibinfo {volume} {116}},\ \bibinfo {pages} {463} (\bibinfo {year} {2018})}\BibitemShut {NoStop}%
\bibitem [{\citenamefont {Sasv{\'a}ri}\ and\ \citenamefont {Urbanovics}(2023)}]{sasvari2023current}%
  \BibitemOpen
  \bibfield  {author} {\bibinfo {author} {\bibfnamefont {P.}~\bibnamefont {Sasv{\'a}ri}}\ and\ \bibinfo {author} {\bibfnamefont {A.}~\bibnamefont {Urbanovics}},\ }in\ \href@noop {} {\emph {\bibinfo {booktitle} {Proceedings of the Central and Eastern European eDem and eGov Days 2023}}}\ (\bibinfo {year} {2023})\ pp.\ \bibinfo {pages} {191--197}\BibitemShut {NoStop}%
\bibitem [{\citenamefont {Csomós}\ and\ \citenamefont {Farkas}(2023)}]{csomos_understanding_2023}%
  \BibitemOpen
  \bibfield  {author} {\bibinfo {author} {\bibfnamefont {G.}~\bibnamefont {Csomós}}\ and\ \bibinfo {author} {\bibfnamefont {J.~Z.}\ \bibnamefont {Farkas}},\ }\href {\doibase 10.1007/s11192-022-04586-1} {\bibfield  {journal} {\bibinfo  {journal} {Scientometrics}\ }\textbf {\bibinfo {volume} {128}},\ \bibinfo {pages} {803} (\bibinfo {year} {2023})}\BibitemShut {NoStop}%
\bibitem [{\citenamefont {Ajzen}(1991)}]{ajzen1991}%
  \BibitemOpen
  \bibfield  {author} {\bibinfo {author} {\bibfnamefont {I.}~\bibnamefont {Ajzen}},\ }\href {\doibase 10.1016/0749-5978(91)90020-T} {\bibfield  {journal} {\bibinfo  {journal} {Organizational Behavior and Human Decision Processes}\ }\textbf {\bibinfo {volume} {50}},\ \bibinfo {pages} {179} (\bibinfo {year} {1991})}\BibitemShut {NoStop}%
\bibitem [{\citenamefont {Moksness}\ \emph {et~al.}(2020)\citenamefont {Moksness}, \citenamefont {Olsen},\ and\ \citenamefont {Tuu}}]{moksness2020}%
  \BibitemOpen
  \bibfield  {author} {\bibinfo {author} {\bibfnamefont {L.}~\bibnamefont {Moksness}}, \bibinfo {author} {\bibfnamefont {S.~O.}\ \bibnamefont {Olsen}}, \ and\ \bibinfo {author} {\bibfnamefont {H.~H.}\ \bibnamefont {Tuu}},\ }\href {\doibase 10.1108/JD-11-2019-0220} {\bibfield  {journal} {\bibinfo  {journal} {Journal of Documentation}\ }\textbf {\bibinfo {volume} {76}},\ \bibinfo {pages} {1393} (\bibinfo {year} {2020})}\BibitemShut {NoStop}%
\bibitem [{\citenamefont {Sverdlichenko}\ \emph {et~al.}(2022)\citenamefont {Sverdlichenko}, \citenamefont {Xie},\ and\ \citenamefont {Margolin}}]{sverdlichenko2022}%
  \BibitemOpen
  \bibfield  {author} {\bibinfo {author} {\bibfnamefont {I.}~\bibnamefont {Sverdlichenko}}, \bibinfo {author} {\bibfnamefont {S.}~\bibnamefont {Xie}}, \ and\ \bibinfo {author} {\bibfnamefont {E.}~\bibnamefont {Margolin}},\ }\href {\doibase 10.1016/j.jemep.2022.100758} {\bibfield  {journal} {\bibinfo  {journal} {Ethics, Medicine and Public Health}\ }\textbf {\bibinfo {volume} {21}},\ \bibinfo {pages} {100758} (\bibinfo {year} {2022})}\BibitemShut {NoStop}%
\bibitem [{\citenamefont {Migheli}\ and\ \citenamefont {Ramello}(2013)}]{migheli2013}%
  \BibitemOpen
  \bibfield  {author} {\bibinfo {author} {\bibfnamefont {M.}~\bibnamefont {Migheli}}\ and\ \bibinfo {author} {\bibfnamefont {G.~B.}\ \bibnamefont {Ramello}},\ }\href {\doibase 10.1007/s10657-013-9388-x} {\bibfield  {journal} {\bibinfo  {journal} {European Journal of Law and Economics}\ }\textbf {\bibinfo {volume} {35}},\ \bibinfo {pages} {149} (\bibinfo {year} {2013})}\BibitemShut {NoStop}%
\bibitem [{\citenamefont {Hakhverdian}\ and\ \citenamefont {Mayne}(2012)}]{hakhverdian2012}%
  \BibitemOpen
  \bibfield  {author} {\bibinfo {author} {\bibfnamefont {A.}~\bibnamefont {Hakhverdian}}\ and\ \bibinfo {author} {\bibfnamefont {Q.}~\bibnamefont {Mayne}},\ }\href {\doibase 10.1017/S0022381612000412} {\bibfield  {journal} {\bibinfo  {journal} {The Journal of Politics}\ }\textbf {\bibinfo {volume} {74}},\ \bibinfo {pages} {739} (\bibinfo {year} {2012})}\BibitemShut {NoStop}%
\bibitem [{\citenamefont {Tan}\ \emph {et~al.}(2017)\citenamefont {Tan}, \citenamefont {Liu}, \citenamefont {Huang},\ and\ \citenamefont {Zheng}}]{tan2017}%
  \BibitemOpen
  \bibfield  {author} {\bibinfo {author} {\bibfnamefont {X.}~\bibnamefont {Tan}}, \bibinfo {author} {\bibfnamefont {L.}~\bibnamefont {Liu}}, \bibinfo {author} {\bibfnamefont {Z.}~\bibnamefont {Huang}}, \ and\ \bibinfo {author} {\bibfnamefont {W.}~\bibnamefont {Zheng}},\ }\href {\doibase 10.1111/pops.12341} {\bibfield  {journal} {\bibinfo  {journal} {Political Psychology}\ }\textbf {\bibinfo {volume} {38}},\ \bibinfo {pages} {469} (\bibinfo {year} {2017})}\BibitemShut {NoStop}%
\bibitem [{\citenamefont {K{\"o}bis}\ \emph {et~al.}(2018)\citenamefont {K{\"o}bis}, \citenamefont {Iragorri-Carter},\ and\ \citenamefont {Starke}}]{kobis2018}%
  \BibitemOpen
  \bibfield  {author} {\bibinfo {author} {\bibfnamefont {N.~C.}\ \bibnamefont {K{\"o}bis}}, \bibinfo {author} {\bibfnamefont {D.}~\bibnamefont {Iragorri-Carter}}, \ and\ \bibinfo {author} {\bibfnamefont {C.}~\bibnamefont {Starke}},\ }in\ \href@noop {} {\emph {\bibinfo {booktitle} {Corruption and norms: Why informal rules matter}}},\ \bibinfo {editor} {edited by\ \bibinfo {editor} {\bibfnamefont {I.}~\bibnamefont {Kubbe}}\ and\ \bibinfo {editor} {\bibfnamefont {A.}~\bibnamefont {Engelbert}}}\ (\bibinfo  {publisher} {Springer},\ \bibinfo {year} {2018})\ pp.\ \bibinfo {pages} {31--52}\BibitemShut {NoStop}%
\end{thebibliography}%
\bibliographystyle{apsrev4-1}

\end{document}